\documentclass[twocolumn]{aastex631}
\shortauthors{Kamiido Kazuki and Yutaka Ohira}
\usepackage{natbib}
\usepackage{physics}
\begin{document}
\title{Collisionless shocks having relativistic velocities in relativistically hot plasmas}
\author[0009-0002-9063-8827]{Kazuki Kamiido}
\author[0000-0002-2387-0151]{Yutaka Ohira}
\affiliation{Department of Earth and Planetary Science, The University of Tokyo, \\
7-3-1 Hongo, Bunkyo-ku, Tokyo 113-0033, Japan}
\email{kamiido-kazuki8990@g.ecc.u-tokyo.ac.jp}
\begin{abstract}
Shocks in relativistically hot plasmas are thought to exist in various high-energy astrophysical phenomena, but it is not clear how relativistic collisionless shocks are formed, whether particles are accelerated by the shock as in the case of cold upstream. 
In this work, collisionless shocks with a relativistic shock velocity in relativistically hot unmagnetized electron-positron plasmas are investigated by 
two-dimensional particle-in-cell simulations. 
It is shown that the upstream flow is dissipated by the Weibel instability, so that the relativistic collisionless shock is formed as in the case of cold upstream. 
The density and magnetic field structures around the shock front are almost independent of the upstream temperature when the spatial scale is normalized by the inertial length scale which takes into account the relativistic temperature. 
This can be understood by considering the pressure anisotropy, which asymptotically approaches a finite value due to the relativistic beaming effect, even as the temperature becomes relativistically hotter and hotter.
In addition, as long as the shock velocity is relativistic, some particles are accelerated, forming a power-law energy spectrum similar to that in the cold upstream. 
\end{abstract}
\section{introduction}
\label{sec1}
Collisionless shocks play an important role in the dissipation of kinetic energy in various astrophysical objects such as the Sun, supernova remnants, galaxy clusters, black holes, and neutron stars.
The dissipation in a collisionless shock converts the upstream kinetic energy to the downstream thermal energy of particles, electromagnetic fields, and nonthermal high energy particles \citep[e.g.][]{quest88,shimada00,spitkovsky08b}.
The some fraction of the thermal and nonthermal particle energy is eventually converted to photons, neutrino, and cosmic ray energies. 

The Weibel mediated collisionless shock has been actively studied by particle-in-cell (PIC) simulations because the simulation time is shorter than that for other types of collisionless shocks.
In the Weibel mediated shock, the upstream plasma is thermalized by magnetic field turbulence generated by the Weibel instability \citep{weibel59}.
This instability arises from anisotropic velocity distributions, leading to the generation of filamentary current structures and magnetic fields.
PIC simulations showed that Weibel-mediated relativistic shocks are formed in unmagnetized plasmas \citep{kato07, spitkovsky08b}, particles are accelerated by the shock \citep{spitkovsky08b, martins09}, and electrons are efficiently heated to near ion temperature \citep{spitkovsky08a}.
The electron heating in the Weibel mediated shock is theoretically well explained \citep{vanthieghem22}.
Recent work by \citet{parsons24} further examined the injection mechanism to the particle acceleration before particles enter the downstream region.
\citet{sironi13} established a critical threshold for Weibel-mediated shock formation, showing that such shocks emerge when the upstream magnetic energy fraction satisfies $B_\mathrm{up}^2 / 4\pi \Gamma_\mathrm{up} n_\mathrm{up} m c^2 \lesssim 10^{-3}$ where $B_\mathrm{up}$, $\Gamma_\mathrm{up}$, $n_\mathrm{up}$, $m$, and $c$ are the upstream magnetic field, the upstream Lorentz factor, the upstream number density at the downstream rest frame, the rest mass of a particle, and the light speed, respectively.
In addition, their study showed that particle acceleration in the Weibel mediated shock is insufficient compared to the prediction in the Bohm limit diffusion.
Moreover, \citet{lemoine19} and \citet{pelletier19} developed a theoretical model describing how the interaction between self-generated electromagnetic fields and incoming plasma streams shapes the shock profile.
These results about the Weibel mediated shock are well summarized in \citet{vanthieghem20}.
Recently, \citet{groselj24} performed a long-term simulation and showed that the typical length scale of the magnetic field turbulence increases with time, so that the decay of the turbulent magnetic field becomes weaker in the downstream region.

Despite the fact that there are many studies about collisionless shocks, ollisionless shocks in relativistically hot plasmas ($T \gg mc^2$, where $T$ is the upstream temperature and we define the Boltzman constant $k_\mathrm{B}=1$) have been rarely investigated. 
However, such collisionless shocks are thought to exist in the high energy astrophysical objects, such as the prompt emission region of gamma-ray bursts \citep{piran04,kumar15}, downstream of relativistic shocks propagating in a inhomogeneous medium \citep{inoue11,tomita22}, and lobes of radio galaxies \citep{matthews19}. 

To understand the dissipation in a collisionless shock in a relativistically hot plasma, as a first step, we investigated collisionless shocks propagating with the nonrelativistic and mildly relativistic velocity in a relativistically hot unmagnetized electron-positron plasma using two-dimensional particle-in-cell (PIC) simulations \citep{kamiido24}. 
Our simulation showed that shocks are mediated by the Weibel instability, the strength of the Weibel magnetic fields becomes smaller as the shock velocity becomes smaller, and particle acceleration by the shock is not observed in the simulation time. 
In \citet{kamiido24}, we discussed the reason for this shock velocity dependence in terms of the dissipation efficiency at the shock front, but did not discuss it in terms of the plasma instability. 
In addition, the temperature dependence and the case with a relativistic shock velocity were not investigated. 
In this study, we investigate collisionless shocks propagating with a relativistic shock velocity in relativistically hot unmagnetized electron-positron plasmas.
The simulation setup and results are presented in Sections \ref{sec2} and \ref{sec3}, respectively. 
To understand the simulation results, we calculate the pressure anisotropy in the shock transition region in Section \ref{sec4}. 
The discussion and summary are given in Section \ref{sec5}.
In the appendices \ref{sec:a}, \ref{sec:b}, and \ref{sec:c}, we derive some physical quantities for the relativistic plasma.

\section{Simulation setup}
\label{sec2}
\begin{figure}[t]
\centering
\includegraphics[scale=0.4]{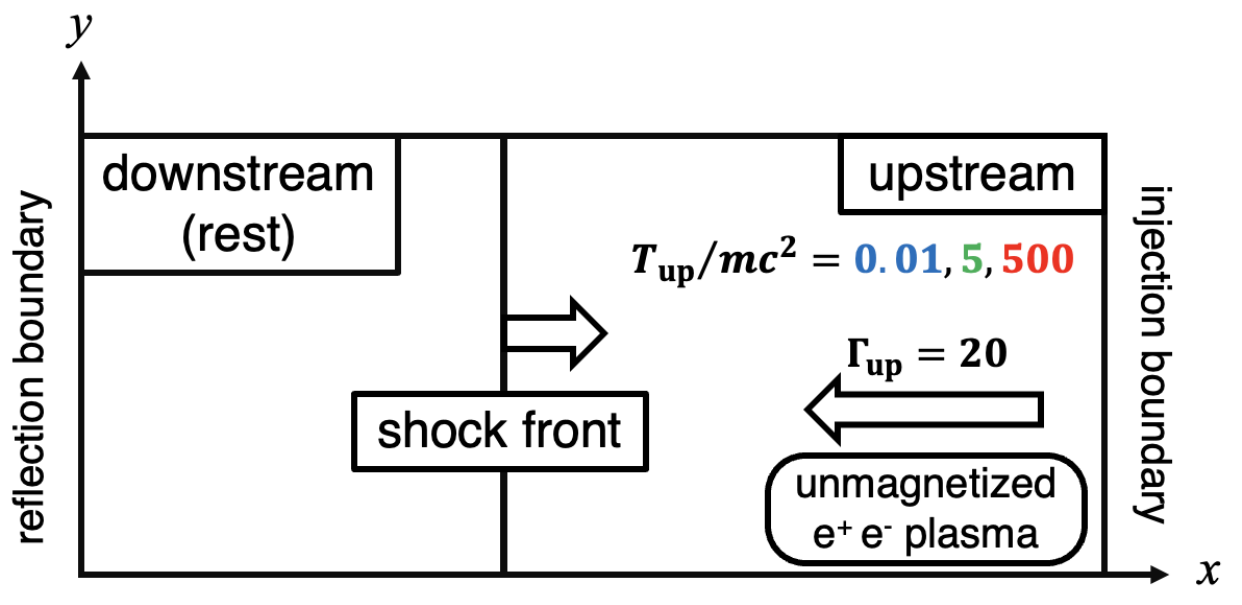}
\caption{Simulation setup. In our simulation, an unmagnetized electron-positron plasma is injected from the injection boundary with the bulk Lorentz factor $\Gamma_\mathrm{up}=20$. The plasma is reflected at the reflection boundary. Once a shock is formed, the shock front propagates to the $+x$ direction. 
The temperature of the injection plasma is set to $T_\mathrm{up}/mc^2=0.01,\,5$, and $500$ for three simulations.
} 
\label{fig:setup}
\end{figure}
2D PIC simulations are performed to investigate collisionless shocks in relativistically hot plasmas.
In PIC simulations, the equations of motion for many charged superparticles and the Maxwell equations are solved selfconsistently.
We use the open code, Wuming \citep{matsumoto24}, which uses the Buneman-Boris method for the equations of motion, the implicit FTDT scheme for the Maxwell equations \citep{hoshino13}, Esirkepov's charge conservation scheme for the current deposition with the second-order shape function \citep{esirkepov01}.
The simulation overview is shown in figure \ref{fig:setup}.
Particles with bulk velocity in the $-x$ direction are continuously injected from the injection boundary and reflected at the left reflection boundary, and if a shock is formed, it propagates to the $+x$ direction. The simulation corresponds to the downstream rest frame.

In this work, the upstream plasma is set to an unmagnetized electron-positron plasma with the Maxwell-J\"{u}ttner distribution. The Maxwell-J\"uttner distribution is generated by the Sobol method \citep{sobol76, zenitani15}. 
Cases of magnetized and electron-ion plasmas will be investigated in future work. 
Three simulations are performed for different temperatures ($T_\mathrm{up}/mc^2 = 0.01, 5,$ and $500$) to see the temperature dependence. 
All other parameters are the same in the three simulations as below. 
The upstream bulk Lorentz factor in the simulation frame (downstream rest frame) is $\Gamma_\mathrm{up}=20$. 
The upstream magnetic field is zero.  
The cell size and time step are set to $\Delta{x}=\Delta{y}=0.05\,c/\omega_\mathrm{p}$ and $\Delta{t}=0.05/\omega_\mathrm{p}$ respectively. $\omega_\mathrm{p}$ is the plasma frequency defined to include the relativistic temperature effect (equation \ref{eq:omega_p}),
\begin{equation}
\omega_\text{p}^2 = \frac{4\pi n_\mathrm{up}e^2}{\langle \gamma \rangle m}
\, .
\label{eq:wpe}
\end{equation}
$n_\mathrm{up}$ is the upstream number density at the downstream rest frame (simulation frame).
$\langle \gamma \rangle$ is the mean Lorentz factor of particles in the simulation frame and given by (Equation \ref{eq:g})
\begin{equation}
\langle \gamma \rangle= \Gamma_\mathrm{up} \qty[\frac{T_\mathrm{up}}{mc^2}\qty(3+\beta_\mathrm{up}^2) + \frac{K_1(mc^2/T_\mathrm{up})}{K_2(mc^2/T_\mathrm{up})}]\, ,
\end{equation}
where $K_n$ is the modified Bessel function of the second kind and $\beta_\mathrm{up}c = c (1-\Gamma_\mathrm{up}^{-2})^{1/2}$ is the the upstream bulk velocity and $n_\mathrm{up}$ is the upstream number density in the simulation frame.
The size of the simulation box is $L_x\times L_y = 4250\,c/{\omega_\mathrm{p}} \times 52 \,c/\omega_\mathrm{p}$. 
$20$ superparticles per cell for electrons and positrons are distributed in the upstream region. 

\section{Result}
\label{sec3}
All figures in this section are simulation results at $t\omega_\mathrm{p}=2000$. 
\subsection{Density distribution}
\begin{figure}[t]
\centering
\includegraphics[scale=0.31]{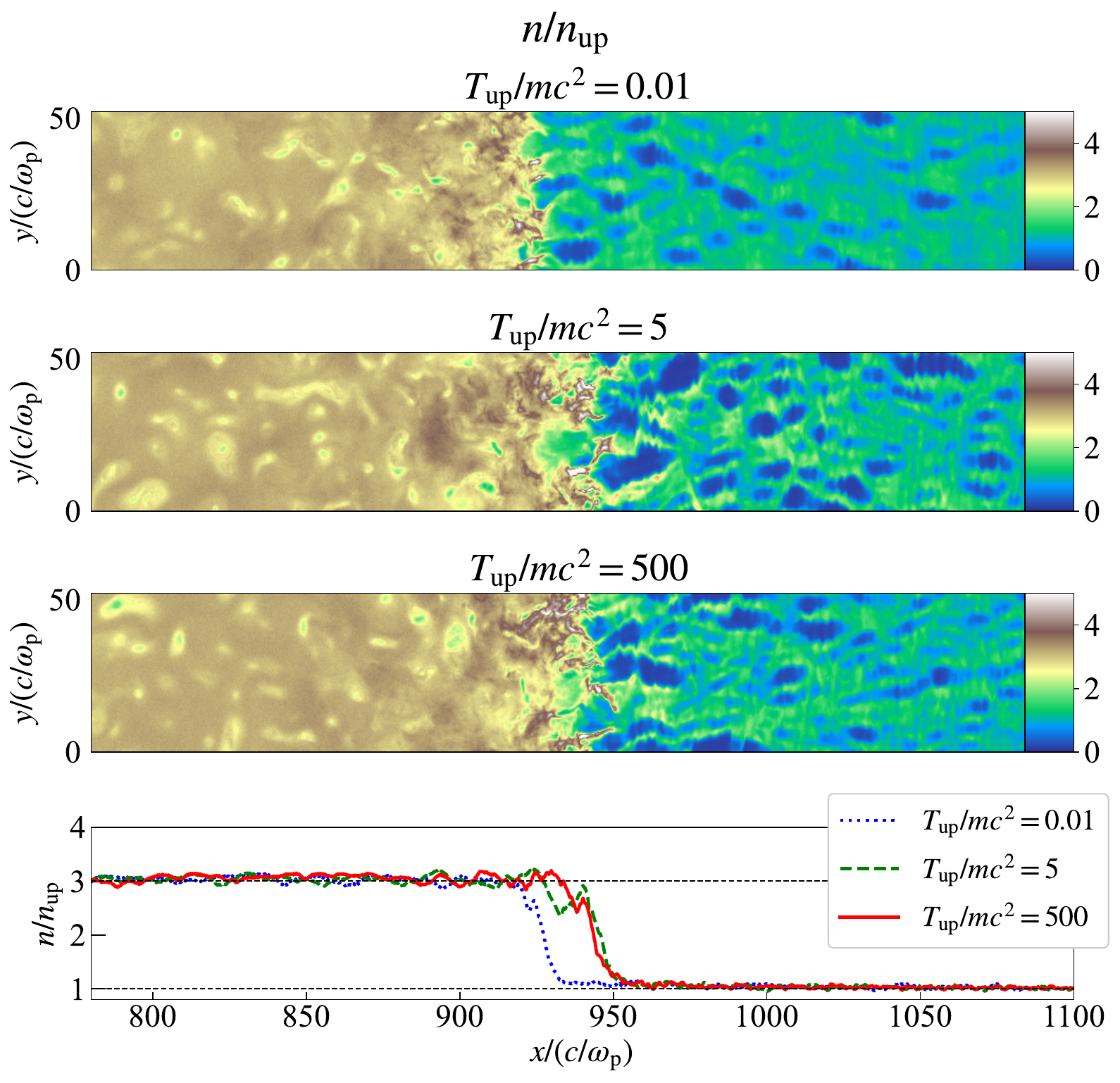}
\caption{Density distribution at $t\omega_\mathrm{p}=2000$.
The density $n$ is normalized by the density at the far upstream, $n_\mathrm{up}$.
The top three panels show the two-dimensional distributions for $T_\mathrm{up}/mc^2=0.01, 5$, and $500$.
The bottom panel shows the one-dimensional distributions averaged over the $y$ direction, where the dotted, dashed, and solid lines show the results for $T_\mathrm{up}/mc^2=0.01, 5$, and $500$, and the black dotted horizontal lines represent $n/n_\mathrm{up}=1$ and $3$.} 
\label{fig:n}
\end{figure}
Figure \ref{fig:n} shows the density distributions normalized by the far upstream density, $n_\mathrm{up}$.
The top three panels are the two-dimensional density distributions for $T_\mathrm{up}/mc^2 = 0.01, 5, 500$, and the bottom panel shows the one-dimensional density distributions averaged over the $y$ direction. 
Note that the spatial scale is normalization by $c/\omega_\mathrm{p}$ which depends on the upstream temperature, $T_\mathrm{up}$ (see Equation (\ref{eq:wpe})).
Our simulations first showed that relativistic collisionless shocks are formed in unmagnetized electron-positron plasmas even though the upstream temperatures are relativistically hot. 
The bottom panel shows that the compression ratio in all simulations are consistent with the Rankine-Hugoniot relation (\ref{eq:n_ratio}) for the two-dimensional case (adiabatic index of $\hat{\gamma}=3/2$). 
The thickness of the shock transition layer is comparable to the size of cavities around the shock front, which is about $10\,c/\omega_\mathrm{p}$ for the nonrelativistic upstream temperature of $T_\mathrm{up}/mc^2=0.01$ and about $20\,c/\omega_\mathrm{p}$ for the relativistic upstream temperature of $T_\mathrm{up}/mc^2=5$ and $500$, respectively. 
Moreover, the contrast of the density fluctuation around the shock front is larger for the shocks in the relativistic upstream temperatures. 

\citet{groselj24} performed a very long-term ($t \omega_\mathrm{p}= 26100$) simulation for the relativistic Weibel mediated collision's shock propagating in the electron-positron plasma with the norelativistic temperature. 
Their long-term simulation showed that the size of the cavity and thickness of the shock transition layer increase with simulation time. 
Even though the temperature is nonrelativistic in the far upstream region, it is expected that, as time goes on, leakage of particles from downstream to upstream heats the upstream plasma around the shock front to the relativistic temperature. 
Thus, our simulation result suggests that heating in the upstream region is one of the reasons for the increase in cavity size observed in \citet{groselj24}.

\begin{figure}[t]
\centering
\includegraphics[scale=0.31]{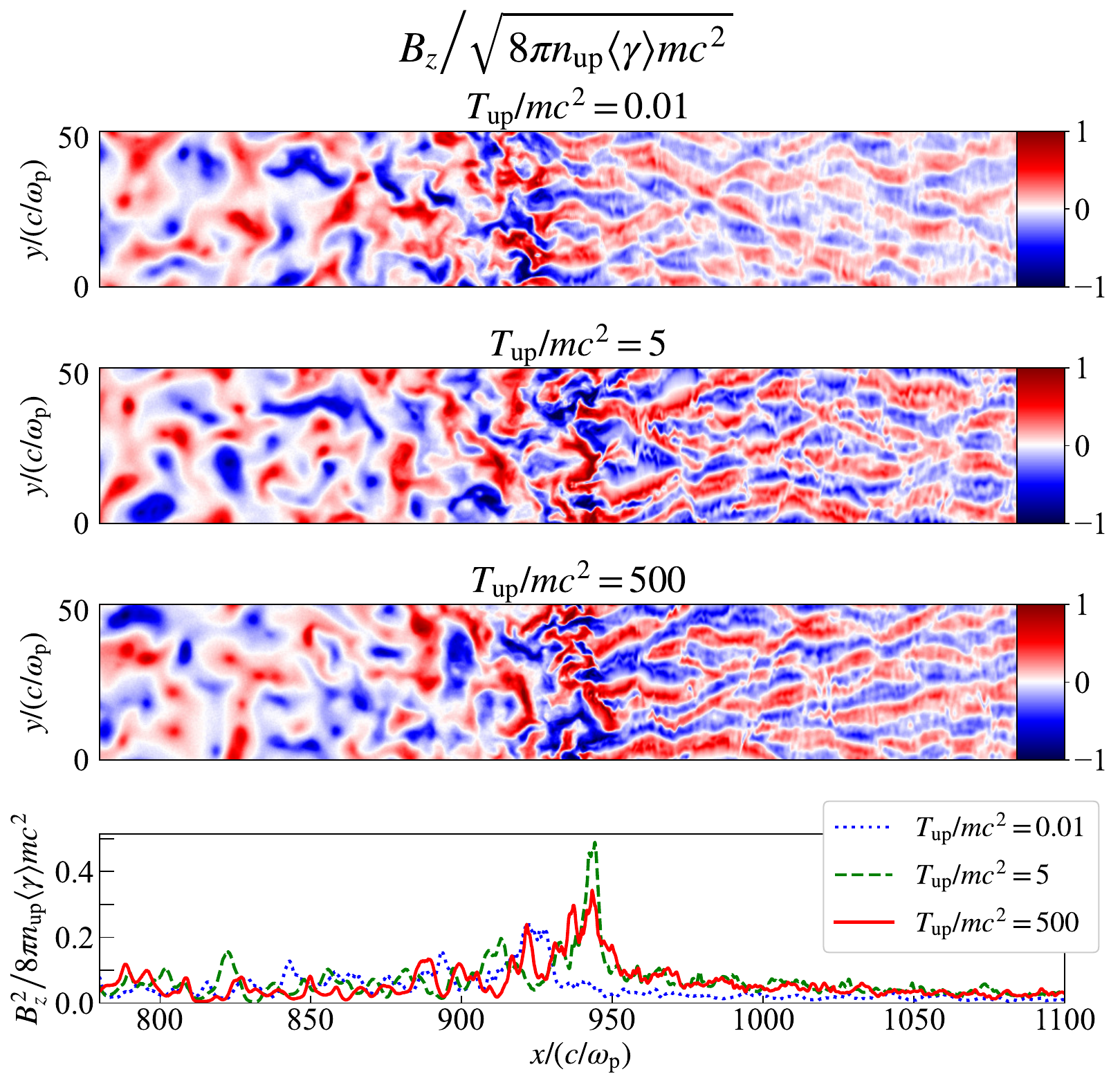}
\caption{The same as Figure~\ref{fig:n}, but for $z$-component of the magnetic field. The magnetic field is normalized by $\sqrt{8\pi n_\mathrm{up}\langle \gamma \rangle mc^2}$}. 
\label{fig:B}
\end{figure}
\subsection{Magnetic field distribution}
Figure \ref{fig:B} shows the distribution of the $z$-component of the magnetic field at $t\omega_\mathrm{p}=2000$, which is normalized by the upstream energy density $n_\mathrm{up}\langle \gamma \rangle mc^2$.
The maximum value and typical wavelength of the magnetic field are $B_z/\sqrt{8\pi n_\mathrm{up}\langle \gamma \rangle mc^2} \sim 0.3$ and $\lambda \sim8\,c/\omega_\mathrm{p}$ in common for all simulations, which are almost identical to the results in the case of cold upstream \citep{spitkovsky08b}.  In addition, the decay rate of the downstream magnetic field is also common to all simulations.
Since the Weibel-like magnetic field structure is observed in all simulations, our simulations show that the relativistic unmagnetized collisionless shock is mediated by the Weibel instability, even though the upstream temperature is relativistically hot.

\subsection{Energy spectrum}
\begin{figure}[t]
\centering
\includegraphics[scale=0.4]{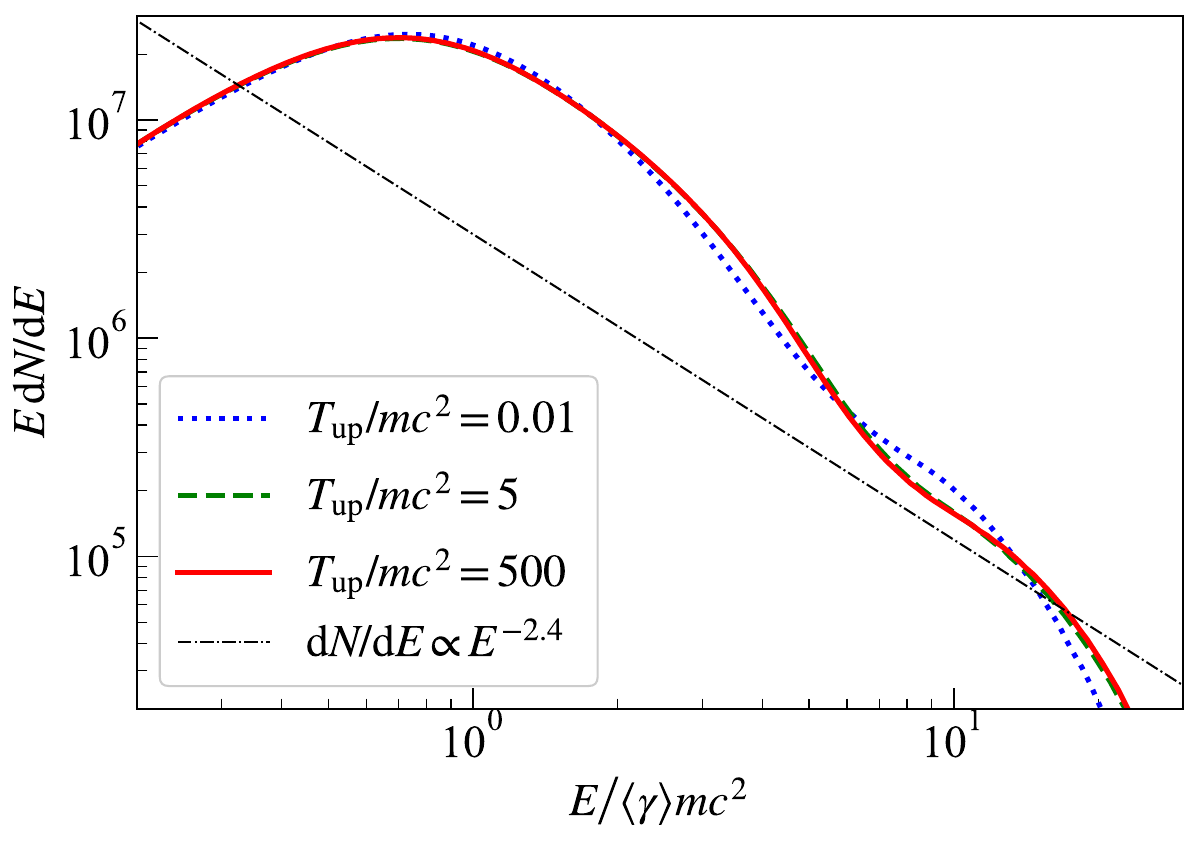}
\caption{Energy spectra in the downstream region ($500<x/(c/\omega_\mathrm{p})<800$) at $t\omega_\mathrm{p}=2000$.
The horizontal axis shows the kinetic energy of particles $E=(\gamma-1)mc^2$ normalized by the mean upstream energy par particle,  $\langle \gamma \rangle mc^2$. 
The black dash-dotted line shows the power law spectrum of $\dd N / \dd E \propto E^{-2.4}$.}
\label{fig:spec}
\end{figure}
Figure \ref{fig:spec} shows the energy spectrum in the downstream region of $500 < x/(c/\omega_\mathrm{p}) < 800$ at $t\omega_\mathrm{p}=2000$. 
As one can see, the energy spectra have the power-law tails with the spectral index of about 2.4 for all simulations, that is, our simulations show that, as in the early studies for the relativistic Weibel mediated shocks in cold plasmas \citep{spitkovsky08b,sironi13}, particles are accelerated by the relativistic unmagnetized collisionless shock in the relativistically hot electron-positron plasma. 

Although the spectral difference in three simulations is small, the spectra of the thermal component have a harder tail in the energy range of $1\lesssim E/\langle \gamma \rangle mc^2\lesssim4$ for the cases where the upstream temperature is relativistically hot ($T_\mathrm{up}/mc^2=5$ and 500). 
The energy spectrum of our simulation for the cold upstream case ($T_\mathrm{up}/mc^2=0.05$) is similar to that in the early phase of the long-term simulation by \citet{groselj24}. 
On the other hand, the energy spectra for the hot upstream cases ($T_\mathrm{up}/mc^2=5$ and $500$) are similar to that in the late phase of the long-term simulation. 
As discussed above, even though the temperature is nonrelativistically cold in the far upstream region, as time goes on, the upstream plasma around the shock front is expected to be heated to the relativistic temperature. 
Therefore, our simulation results suggest that heating in the upstream region may be responsible for the long-term evolution observed in \citet{groselj24}, such as changes in clump size and energy spectra.

\section{Pressure anisotropy}
\label{sec4}
\begin{figure}[t]
\centering
\includegraphics[scale=0.5]{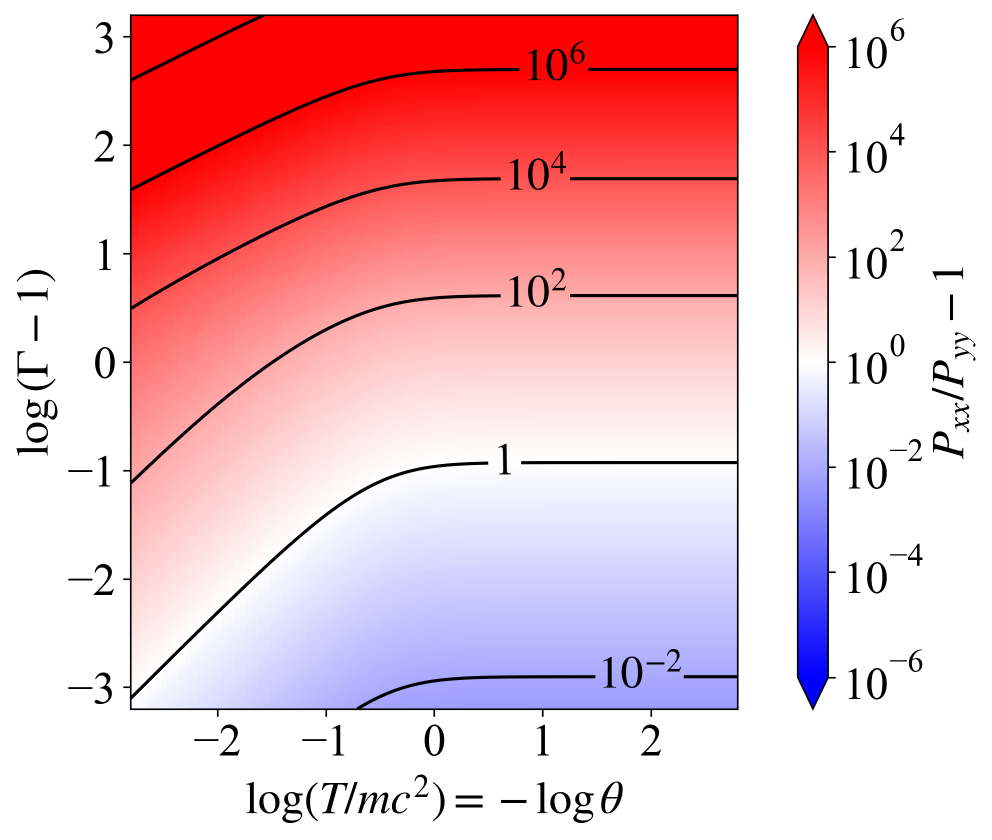}
\caption{Pressure anisotropy, $P_{xx}/P_{yy}-1$ of the counter streaming plasmas with the bulk Lorentz factor, $\Gamma$, and the temperature $T/mc^2=1/\theta$. 
The contours are plotted for $P_{xx}/P_{yy}-1 = 10^{-2}, 1, 10^2, 10^4, 10^6$, and $10^8$, respectively.
} 
\label{fig:aniso}
\end{figure}

As shown in the previous section, even in a relativistically hot unmagnetized plasma, a relativistic collisionless shock is formed by the Weibel instability \citep{weibel59}. 
Furthermore, even though the mean Lorentz factor correspond to the upstream thermal velocity is larger than the bulk Lorentz factor of the upstream flow ($T_\mathrm{up}/mc^2 \gg \Gamma_\mathrm{up}$), the shock structure does not depend on the upstream temperature when physical quantities are normalized appropriately. 
In the nonrelativistic picture, the condition of $T_\mathrm{up}/mc^2 \gg \Gamma_\mathrm{up}$ corresponds to the low Mach number or the low pressure anisotropy in the shock transition region.  
If this picture were also valid for the relativistic case, one would expect the growth rate of the Weibel instability to be smaller, resulting in a wider shock transition layer, which is incompatible with the simulation results in this work. 

In our previous work \citep{kamiido24}, we showed that most of the upstream energy is thermalized in the shock transition region as long as the shock velocity is relativistic even though the upstream temperature is extremely high. In order to thermalize the upstream kientic energy in the collisionless shock, some kinetic plasma instabilities have to develop in the shock transition region.
Since the pressure anisotropy is the most important quantity for the Weibel instability \citep{davidson72, yoon07},  
to understand the above discrepancy, we estimate the pressure anisotropy in the shock transition region by simplifying the system. 

The plasma in the shock transition region consists mainly of two components: the upstream plasma moving from upstream to downstream and the plasma leaking from downstream to upstream. 
As a simple system, we consider counter streaming plasmas with the bulk velocities in the $x$ direction $\pm \beta c$, and the same density and temperature, $n$ and $T$. 
Even for a realistic asymmetric beam system, the pressure anisotropy does not change significantly (see appendix \ref{sec:d} for details). 
%

The momentum distribution of the counter streaming plasmas is assumed to be the shifted Maxwell-J\"{u}ttner distribution function (Equation \ref{eq:f_c}). 
Then, the pressure anisotropy of the counter streaming plasmas is given by (Equation \ref{eq:aniso}),
\begin{align}
\frac{P_{xx}}{P_{yy}} - 1 =
\Gamma^2 \qty[1 + 3\beta^2 + \beta^2 \theta \frac{K_1(\theta)}{K_2(\theta)}] - 1
\, ,
\label{eq:P_aniso}
\end{align}
where $\Gamma = 1/\sqrt{1-\beta^2}$ and $\theta=mc^2/T$, and $P_{xx}$ and $P_{yy}$ are the pressure in the $x$ and $y$ directions, respectively. 
Figure \ref{fig:aniso} shows the pressure anisotropy as a function of $\Gamma-1$ and $T$. 
As the temperature increases while fixing the bulk Lorenz factor, the anisotropy becomes smaller in the nonrelativistic temperature regime ($T\ll mc^2$), but approaches a constant in the relativistic temperature regime ($T\gg mc^2$). 
For $\Gamma-1\gtrsim 0.1$, the pressure anisotropy is larger than unity ($P_{xx}/P_{yy}-1>1$), regardless of the temperature.

\begin{figure}[t]
\centering
\includegraphics[scale=0.4]{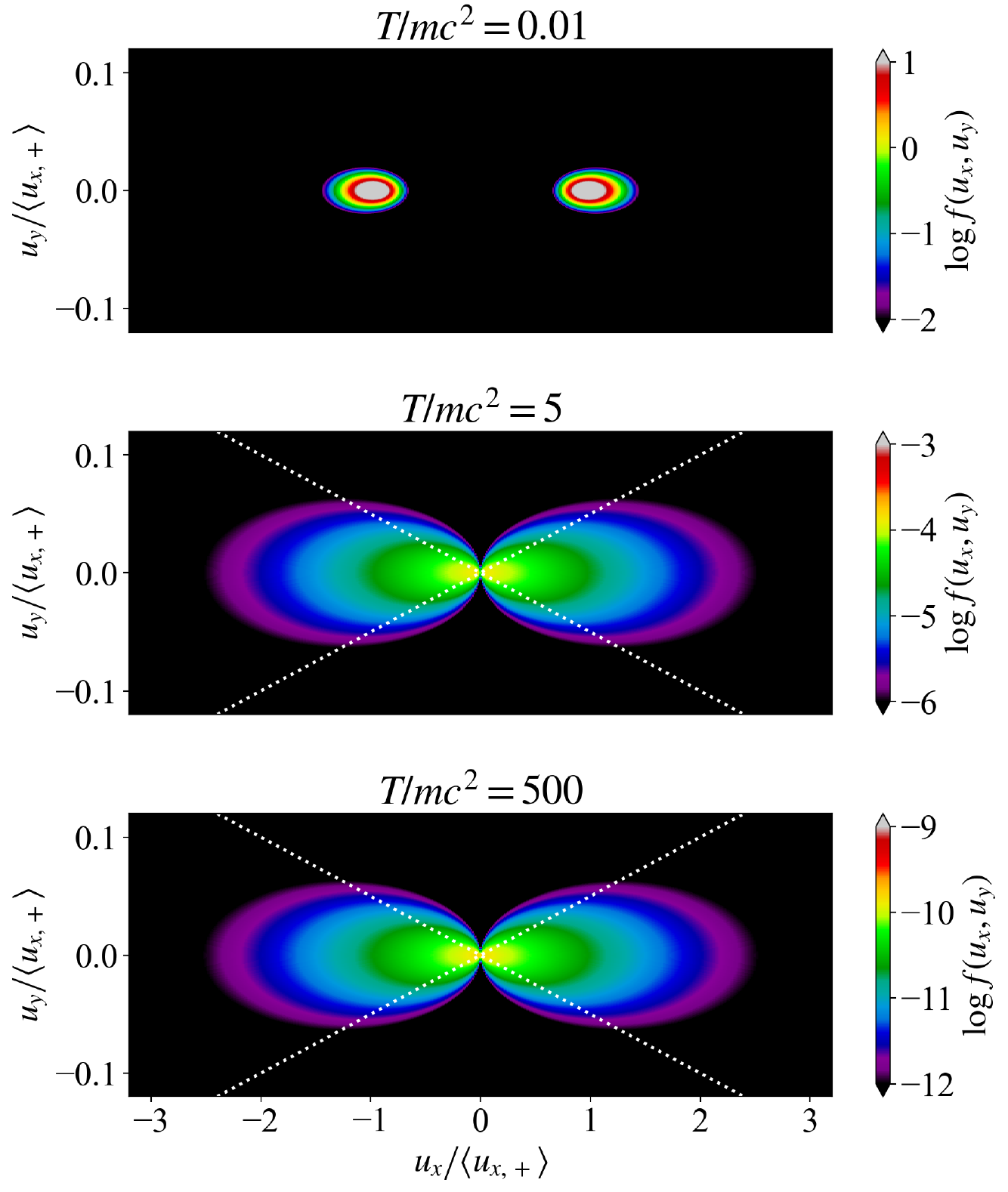}
\caption{Distribution of counter streaming plasmas in the four velocity space. 
The four velocities $\vb{u}$ are normalized by the mean value of the momentum of particles in the beam moving in $+x$ direction, $\langle u_{x,+} \rangle$.} The white lines show the cone of the relativistic beaming effect,  $u_y/u_x = \pm 1/\Gamma$, where $\Gamma = 20$ is the bulk Lorentz factor corresponding to the counter streaming velocity.
\label{fig:momentum}
\end{figure}

Figure \ref{fig:momentum} shows the momentum distribution in the $u_x-u_y$ plane of the counter streaming plasmas with $\Gamma=20$ and $T/mc^2=0.01,5,500$, where $u_x$ and $u_y$ are the four velocity in the $x$ and $y$ directions, and normalized by $\langle u_{x,+} \rangle$.
$\langle u_{x, +} \rangle$ is the mean value of the momentum of particles, $u_x$, in the beam moving in $+x$ direction in this system (equation \ref{eq:ux}).
As shown in the middle and bottom panels in Figure \ref{fig:momentum}, 
the normalized momentum distribution does not depend on the temperature for the relativistic temperature. 
Since almost all particles have velocities very close to the speed of light in the each plasma rest frame for the relativistic temperature, in the counter streaming frame, their angular distribution is concentrated in the flow direction in the angle of $\Gamma^{-1}$. 
This is the so-called relativistic beaming effect. 
Then, the angular distribution of the counter streaming plasmas is almost independent of the temperature. 
This is the reason why the pressure anisotropy does not depend on the relativistic temperature. 

As long as the counter streaming velocity is relativistic ($\Gamma \gtrsim 1.1$), the pressure anisotropy is larger than $1$ even though the temperature is extremely high ($T \gg mc^2$). 
Then, plasma instabilities excited by the pressure anisotropy, such as the Weibel instability, are expected to occur as in cold plasmas. 
Actually, as shown in Figures \ref{fig:n} and \ref{fig:B}, the structure of relativistic shocks mediated by the Weibel instability is almost independent of the upstream temperature. 
\section{Discussion and Summary}
\label{sec5}
In this study, we have performed PIC simulations of collisionless shocks with a relativistic shock velocity, $\Gamma_\mathrm{up}=20$, in unmagnetized relativistically hot electron-positron plasmas. 
The Weibel-mediated relativistic collisionless shock is formed in the simulation even though the upstream temperature is relativistically hot ($T/mc^2 \gg 1$). 
The shock structure is almost independent of the upstream temperature when the spatial scale is normalized appropriately. 
In addition, the efficiency of particle acceleration also does not depend significantly on the upstream temperature, although the energy spectra are slightly different for nonrelativistic and relativistic temperatures. 
These properties can be understood by considering the pressure anisotropy in the shock transition region. 
Even though the upstream temperature is relativistically high ($T/mc^2 \gg 1$), the pressure anisotropy has a sufficiently large value in the shock transition region due to the relativistic beaming effect. 
At a relativistic shock velocity, due to this relativistic beaming effect, electromagnetic fields are generated through the plasma instabilities, so that a collisionless shock is formed even as the upstream temperature becomes relativistically hotter and hotter. 

When a plasma collides at a supersonic speed, a shock wave is always created in the hydrodynamics and magnetohydrodynamics. 
Kinetic plasma simulations are necessary to understand the time and spatial scales required for the shock formation in the collisionless plasma system. 
This work has showed that collisionless shocks are formed on the spatial and time scales of the kinetic plasma system, $\sim c/\omega_\mathrm{p}$ and $\sim \omega_\mathrm{p}^{-1}$, even in the relativisitcally hot unmagnetized plasma, which are much smaller than astrophysical scales.

Shocks in relativistically hot plasmas are expected to exist in various astrophysical phenomena such as gamma-ray bursts, active galactic nuclei, and pulsar winds. 
Our simulation results indicate that such shocks can be formed as collisionless shocks, and the magnetic field generations and particle accelerations occur as in the cold upstream case. 
Therefore, high-energy cosmic rays, neutrinos, and gamma rays are expected to be generated not only shocks having cold upstreams but also shocks having relativistically hot upstreams.

When a relativistic shock propagates in a cold inhomogeneous density region, secondary shocks are  formed in the relativistically hot downstream of the main shock \citep{inoue11,tomita22}.
In this case, the main shock in the cold plasma accelerates particles and amplifies magnetic fields by the turbulence \citep{morikawa24}. 
Then, the upstream plasma of the secondary shock has the magnetic turbulence and a nonthermal energy spectrum rather than the Maxwell-J\"{u}ttner distribution. 
Therefore, collisionless shocks in magnetized relativistically hot plasmas and reacceleration by the secondary shock in the downstream region of the main shock should be investigated in the future.

%

%

\begin{acknowledgments}
We thank T. Amano, and T. Jikei for helpfull discussions.
Numerical computations were carried out on Cray XC50 at the Center for Computational Astrophysics, National Astronomical Observatory of Japan. 
K.K. is supported by International Graduate Program for Excellence in Earth-Space Science (IGPEES). 
Y.O. is supported by JSPS KAKENHI grants No. JP21H04487 and No. JP24H01805.
\end{acknowledgments}
\appendix
\twocolumngrid
\section{Plasma with a relativistic temperature}
\label{sec:a}
The thermal distribution of relativistically hot plasma moving in the $x$ direction is given by the shifted Maxwell-J\"{u}ttner distribution function \citep{juttner11}, 
\begin{align}
\label{eq:Juttner}
f(\vb{u}) = \frac{n \theta}{4\pi K_2(\theta)} e^{-\Gamma\theta(\gamma - \beta u_x)} \, ,
\end{align}
where $\theta = mc^2/T$, and $\gamma$ and $\vb{u}$ are the particle Lorentz factor and particle four velocity normalized by the speed of light ($\gamma=\sqrt{1+u^2}$), and $\Gamma$, $\beta$, and $n$ are the bulk Lorentz factor, bulk velocity, and number density of the plasma in the laboratory frame  ($\Gamma=1/\sqrt{1-\beta^2}$), respectively. 
$K_n(\theta)$ is the modified Bessel function of the second kind. 

\begin{figure}[t]
\centering
\includegraphics[scale=0.50]{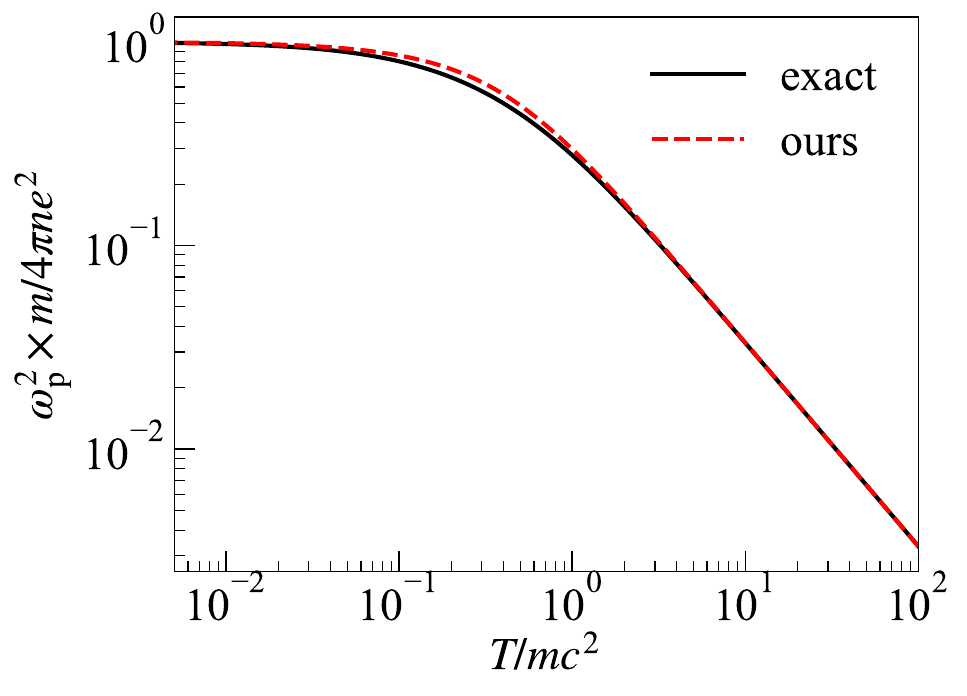}
\caption{The exact plasma frequency in \citep{bergman01} (black solid line) and our definition (red dashed line).} 
\label{fig:omega_p}
\end{figure}

Then, the mean Lorentz factor of particles in the laboratory frame is given by 
\begin{eqnarray}
\label{eq:g}
\langle \gamma \rangle 
&=& 
\frac{1}{n}\int\dd^3 \, 
u \gamma f(\vb{u}), 
\nonumber \\
&=& 
\Gamma \qty[\frac{1}{\theta}\qty(3+\beta^2) + \frac{K_1(\theta)}{K_2(\theta)}] . 
\end{eqnarray}
The fraction of $K_1(\theta)/K_2(\theta)$ can be approximated as 
$K_1(\theta)/K_2(\theta)\rightarrow1 - 3/2\theta$ in the cold limit and $K_1(\theta)/K_2(\theta)\rightarrow\theta/2$ in the relativistically hot limit.  
Using the mean Lorentz factor of particles equation (\ref{eq:g}), the plasma frequency $\omega_\text{p}$ is given by
\begin{eqnarray}
\label{eq:omega_p}
\omega_\text{p}^2 &=& \frac{4\pi ne^2}{\langle \gamma \rangle m} \nonumber \\
&=& \frac{4\pi ne^2}{\Gamma m} \qty[\frac{1}{\theta}\qty(3+\beta^2) + \frac{K_1(\theta)}{K_2(\theta)}]^{-1}\, .
\end{eqnarray}
where $e$ and $m$ are the charge and mass of each particle. 
The exact expression of the plasma frequency in the plasma rest frame is given by the dispersion relation in the Vlasov equation system \citep{bergman01},
\begin{align}
\omega_\mathrm{p}^2 = 
G^{30}_{10} \left . \qty(\left. \frac{\theta^2}{4} \right|^2_{-1/2,1,3/2}) \right / 2 K_2(\theta),
\end{align}
where $G$ is the Meijer's G function. Our definition of the plasma frequency is easily to understand due to using the effective mass and there is almost no difference between our definition and the exact expression in the plasma rest frame as shown in figure \ref{fig:omega_p}.

The mean 4-velocity, $\langle u_x \rangle$ in the laboratory frame is also given by
\begin{align}
\label{eq:ux}
\langle u_x \rangle &= \frac{1}{n} \int \dd^3 u \, u_x f(\vb{u})
\nonumber \\
&= \Gamma \beta \qty[\frac{4}{\theta} + \frac{K_1(\theta)}{K_2(\theta)}]\, .
\end{align}
\section{Pressure anisotropy of counter streaming plasmas}
\label{sec:b}
We here derive the equation of the pressure anisotropy in the shock transition region (\ref{eq:aniso}).
Let us consider counter streaming plasmas with the opposite bulk velocities in the $x$ direction, $\pm \beta c$, the same bulk Lorentz factor, $\Gamma=1/\sqrt{1-\beta^2}$, the same temperature, $T=mc^2/\theta$, and the same number density, $n/2$. 
The pressure tensor, $P_{ij}$, is defined by
\begin{align}
\label{eq:P_definition}
P_{ij} = mc^2 \int \dd^3 u \,
\frac{u_iu_j}{\gamma}f(\vb{u})
\, ,
\end{align}
where $m$, $c$, $\vb{u}$, $\gamma$, and $f(\vb{u})$ are the particle mass, the light speed, the particle four velocity normalized by $c$, the particle Lorentz factor, and the distribution function.
The distribution function of the counter streaming plasmas is given by 
\begin{align}
\label{eq:f_c}
f(\vb{u}) = 
\frac{n\theta}{8 \pi K_2(\theta)}
\qty[e^{-\Gamma\theta(\gamma - \beta u_x)} + e^{-\Gamma\theta(\gamma + \beta u_x) } ]
\, .
\end{align}
Using this distribution function, the pressure anisotropy, $P_{xx}/P_{yy} - 1$, is 
\begin{align}
\label{eq:aniso}
\frac{P_{xx}}{P_{yy}} - 1 =
\Gamma^2 \qty[1 + 3\beta^2 + \beta^2\theta\frac{K_1(\theta)}{K_2(\theta)}] - 1
\, .
\end{align}
\section{Rankine-Hugoniot relation}
\label{sec:c}
The conservation of mass, energy, and momentum fluxes and the equation of state can be written as
\begin{align}
\Gamma_{1}\rho_{1}\beta_{1} &= \Gamma_{2}\rho_{2}\beta_{2} \, 
, \\
\Gamma_{1}^2 (\rho_{1}c^2 + e_{1} +& p_{1})\beta_{1}
\nonumber\\
= \Gamma_{2}^2  ( & \rho_{2}c^2+e_{2}+p_{2})\beta_{2}
\, 
, \\
\Gamma_{1}^2(\rho_{1}c^2+e_{1}+p_{1})\beta_{1}^2 & + p_{1}
\nonumber\\ = \Gamma_{2}^2(\rho_{2}c^2 & + e_{2} + p_{2})\beta_{2}^2 + p_{2}
\, , \\
p_i &= (\hat{\gamma}_i-1)e_i \ \ (i = 1,2) \, ,
\end{align}
where $\beta_i$, $\Gamma_i$, $\rho_i$, $e_i$, $p_i$, $\hat{\gamma}_i$, and $c$ are the bulk three velocity, Lorentz factor of the bulk velocity, mass density, thermal energy density, thermal pressure, adiabatic index, and light speed respectively.
The mass density and internal energy density are defined in the fluid rest frame and the index $i$ means the upstream or downstream ($1$ : upstream, $2$ : downstream).
Using these equations, the following relationships are obtained.
\begin{align}
\beta_{1}^2 &= \frac{(p_{2}-p_{1})(\rho_{2}c^2+e_{2}+p_{1})}{(\rho_{2}c^2+e_{2}-\rho_{1}c^2-e_{1})(\rho_{1}c^2+e_{1}+p_{2})}\, ,
\\
\beta_{2}^2 &= \frac{(p_{2}-p_{1})(\rho_{1}c^2+e_{1}+p_{2})}{(\rho_{2}c^2+e_{2}-\rho_{1}c^2-e_{1})(\rho_{2}c^2+e_{2}+p_{1})}
\, , \\
\Gamma_{1}^2 &= \frac{(\rho_{2}c^2+e_{2}-\rho_{1}c^2-e_{1})}{(\rho_{2}c^2+e_{2}-\rho_{1}c^2-e_{1}-p_{2}+p_{1})}
\nonumber \\ &~~~~~~~~~~~~~~~~~~~~~~~~
\times \frac{(\rho_{1}c^2+e_{1}+p_{2})}{(\rho_{1}c^2+e_{1}+p_{1})}
\, , \\
\Gamma_{2}^2 &= \frac{(\rho_{2}c^2+e_{2}-\rho_{1}c^2-e_{1})}{(\rho_{2}c^2+e_{2}-\rho_{11}c^2-e_{1}-p_{2}+p_{1})} 
\nonumber \\ &~~~~~~~~~~~~~~~~~~~~~~~~
\times \frac{(\rho_{2}c^2+e_{2}+p_{1})}{(\rho_{2}c^2+e_{2}+p_{2})}\, ,
\end{align}
and 
\begin{align}
\beta_{12} &= \frac{\beta_1 - \beta_2}{1 - \beta_1\beta_2} 
\, 
, \\
\Gamma_{12}^2 &= 
\frac{(\rho_{1}c^2+e_{1}+p_{2})(\rho_{2}c^2+e_{2}+p_{1})}{(\rho_{1}c^2+e_{1}+p_{1})(\rho_{2}c^2+e_{2}+p_{2})}
\, ,
\end{align}
where $\beta_{12}$ and $\Gamma_{12} = 1/\sqrt{1-\beta_{12}^2}$ are the upstream three velocity in the downstream rest frame and the Lorentz factor of $\beta_{12}$. 
From the mass conservation and the above relationships, the density ratio is given by
\begin{align}
\qty(\frac{\rho_{1}}{\rho_{2}})^2 = \frac{(\rho_{1}c^2+e_{1}+p_{1})(\rho_{1}c^2+e_{1}+p_{2})}{(\rho_{2}c^2+e_{2}+p_{1})(\rho_{2}c^2+e_{2}+p_{2})}
\, .
\end{align}

For the cold upstream limit ($\rho_1c^2 \gg e_1,p_1$), the downstream quantities are
\begin{align}
\frac{\rho_2}{\rho_1} &=
\frac{\hat{\gamma}_2 \Gamma_{12}+1}{\hat{\gamma}_2-1} 
\, , \\
\frac{\beta_2}{\beta_1} &=
\frac{(\hat{\gamma}_2-1)(\hat{\gamma}_2\Gamma_{12}-\hat{\gamma}_2+1)}{\hat{\gamma}_2\Gamma_{12}+1}
\, , \\
\frac{e_2}{\rho_2 c^2} &=  \Gamma_{12} - 1 
\, .
\end{align}

For the relativistically hot limit, the relation of $\hat{\gamma}_1 = \hat{\gamma}_2 = \hat{\gamma}$ is achieved, and the downstream quantities are
\begin{align}
\beta_1\beta_2 &=
\hat{\gamma} - 1
\, , \\
\frac{e_2}{e_1} &=
\Gamma_1^2\frac{\beta_1^2 - (\hat{\gamma}-1)^2}{\hat{\gamma}-1}
\, , \\
\frac{\rho_2}{\rho_1} 
&= \Gamma_1\beta_1
\frac{\sqrt{\beta_1^2 - (\hat{\gamma}-1)^2}}{\hat{\gamma}-1}
\, .
\end{align}

\,

In the relativistic velocity limit ($\beta_1\rightarrow1$), the density ratio for both the cold and the hot limits is
\begin{align}
\label{eq:n_ratio}
\frac{\rho_2}{\Gamma_{12}\rho_1} =
\frac{\hat{\gamma_2}}{\hat{\gamma_2}-1}
\, .
\end{align}
\section{Center of momentum system of two assymmetric beams}
\label{sec:e}
In appendix \ref{sec:d}, we consider the pressure anisotropy of two assymetric beams.
Because the pressure is defined in the center of momentum system, the three velocities of two beams in the center of momentum system and their Lorentz factors are required.

First of all, we consider two beams ($i=1,2$) having three velocity in the laboratory frame $\beta_i$ in the $+x$ direction, velocity Lorentz factor $\Gamma_i$, the density in the beam rest frame $n^\prime_i$, and the temperature $T_i = mc^2 / \theta_i$.
In the laboratory frame, the center of momentum system moves with the three velocity $\beta_\mathrm{G}$ and the Lorentz factor $\Gamma_\mathrm{G}$.
In the center of momentum system, two beams move with $\beta_{i\mathrm{G}}$ and $\Gamma_{i\mathrm{G}}$.
Using the equation (\ref{eq:ux}), in the center of momentum system, where the total momentum is zero, the equation holds as
\begin{align}
\Gamma_{1\mathrm{G}}^2 \beta_{1\mathrm{G}}
=
r \Gamma_{2\mathrm{G}}^2 \beta_{2\mathrm{G}} \, ,
\end{align}
where
\begin{align}
r = \frac{n_2^\prime}{n_1^\prime}
\frac{4/\theta_2 + K_1(\theta_2)/K_2(\theta_2)}{4/\theta_1 + K_1(\theta_1)/K_2(\theta_1)} \, .
\end{align}

The Lorentz transformations of the three velocity and the Lorentz factor are
\begin{align}
\beta_{i\mathrm{G}} 
&=
\frac{\beta_i - \beta_\mathrm{G}}{1 - \beta_i\beta_\mathrm{G}}\, ,
\\
\Gamma_{i\mathrm{G}} &= \Gamma_i\Gamma_\mathrm{G} (1 - \beta_i\beta_\mathrm{G})\,.
\end{align}
Using these equations, $\beta_\mathrm{G}$ and $\Gamma_\mathrm{G}$ are derived as
\begin{align}
\beta_\mathrm{G} &=
\frac{b - \sqrt{b^2 - 4a^2}}{2a} \, ,
\\
\Gamma_\mathrm{G}^2 &=
\frac{b+\sqrt{b^2-4a^2}}{2 \sqrt{b^2-4a^2}}
\, ,
\end{align}
where
\begin{align}
a &=
\Gamma_1^2\beta_1 + r\Gamma_2^2\beta_2
\, ,\\
b &=
\Gamma_1^2 (1+\beta^2_1) + r \Gamma_2^2 (1+\beta^2_2)
\, , \\
b^2-4a^2 &=
(r+1)^2 + 4r\Gamma_1^2\Gamma_2^2(\beta_1-\beta_2)^2
\, .
\end{align}

Finally, the expression of quantities in the center of momentum system is derived as
\begin{align}
\label{eq:eccentric_b}
\beta_{i\mathrm{G}}
&=
\frac{2a\beta_i - (b-\sqrt{b^2-4a^2})}{2a - \beta_i (b-\sqrt{b^2-4a^2})}
\, , \\
\label{eq:eccentric_g}
\Gamma_{i\mathrm{G}}
&=
\Gamma_i \qty(\frac{b + \sqrt{b^2-4a^2}}{2\sqrt{b^2-4a^2}}) ^{1/2}
\qty(1- \beta_i\frac{b-\sqrt{b^2-4a^2}}{2a})
\end{align}
These become the simple expression when the extreme cases.
When $r \rightarrow 0$, $\Gamma_{1\mathrm{G}} = 1$, and $\Gamma_{2\mathrm{G}} = \Gamma_2\Gamma_1(1-\beta_2\beta_1)$.
When $\beta_2=-\beta_1=-1$ and $r^{1/2}, \, r^{-1/2} \ll \Gamma_1^2\Gamma_2^2$, $\Gamma_{1\mathrm{G}} = r^{1/4}\sqrt{\Gamma_1\Gamma_2}$ and $\Gamma_{2\mathrm{G}} = r^{-1/4}\sqrt{\Gamma_1\Gamma_2}$.
\section{Pressure anisotropy of two asymmetric beams}
\label{sec:d}
In section \ref{sec4}, we discuss the pressure anisotropy in the transition layer using the very simple symmetric system, i.e. counter streaming plasmas.
After time development, the momentum distribution in the shock transition layer becomes asymmetric since there are two components, the dense upstream component and the leakage component from the downstream.
In this section, we show such the asymmetric system is essentially the same as the symmetric system as shown in section \ref{sec4}, and our discussion using the symmetric system is suffice.

We consider two beams in the laboratory frame.
The momentum of two beams having the number density in the beam rest frame $n^\prime_i$, the 3-velocity $\beta_i$, the Lorentz factor $\Gamma_i$, and the temperature $T_i = mc^2 / \theta_i$.
Because the pressure is defined in the center of momentum system, we have to move to the center of momentum system.
As calculated in appendix \ref{sec:b}, 
\begin{align}
P_{xx} &=
\Gamma_{1\mathrm{G}} n_1^\prime 
\qty[\frac{1}{\theta_1}(1+3\beta_{1\mathrm{G}}^2) + \beta_{1\mathrm{G}}^2 \frac{K_1(\theta_1)}{K_2(\theta_1)}]
\nonumber \\
& ~~~ + 
\Gamma_{2\mathrm{G}} n_2^\prime 
\qty[\frac{1}{\theta_2}(1+3\beta_{2\mathrm{G}}^2) + \beta_{2\mathrm{G}}^2 \frac{K_1(\theta_2)}{K_2(\theta_2)}]
\, ,
\end{align}
and
\begin{align}
P_{yy} = 
\frac{n_1^\prime}{\theta_1} + \frac{n_2^\prime}{\theta_2}
\, .
\end{align}
$\beta_{i\mathrm{G}}$ and $\Gamma_{i\mathrm{G}}$ is calculated using equations (\ref{eq:eccentric_b}), (\ref{eq:eccentric_g}).
When we assume $n_1^\prime = n_2^\prime$, $\beta_1 = -\beta_2$, and $\theta_1 = \theta_2$, we can get the same pressure anisotropy in appendix \ref{sec:b}.

In the shock transition layer, two components exists (index $\mathrm{u}$ : upstream fluid and index $\mathrm{l}$ : leakage from the downstream).
Considering the extreme case, $\beta_\mathrm{l} \sim 1$,  $\beta_\mathrm{u} \sim -1$, $\Gamma_\mathrm{u}=\Gamma_\mathrm{l}=\Gamma$, $\theta_\mathrm{u}=\theta_\mathrm{l}=\theta$, and $n_\mathrm{u}^\prime/n_\mathrm{l}^\prime, \, n_\mathrm{l}^\prime/n_\mathrm{u}^\prime \ll \Gamma^8$, the pressure anisotropy is calculated as
\begin{align}
\frac{P_{xx}}{P_{yy}} - 1
=
\Gamma^2 \frac{2\sqrt{n_\mathrm{l}^\prime/ n_\mathrm{u}^\prime}}{1 + n_\mathrm{l}^\prime / n_\mathrm{u}^\prime }
\qty[ 4 + \theta \frac{K_1(\theta)}{K_2(\theta)}] - 1
\, .
\end{align}
As long as $n_\mathrm{l}^\prime / n_\mathrm{u}^\prime \sim \order{0.1}$, the above estimation does not significantly differ from one for the symmetric beam system as shown in section \ref{sec4}.



%

%
\end{document}